\documentstyle[eqsecnum,preprint,aps,prd,epsf,rotate]{revtex}
\date{\today}
\draft
\tighten
\begin{document}
\def\sqr#1#2{{\vcenter{\hrule height.3pt
      \hbox{\vrule width.3pt height#2pt  \kern#1pt
         \vrule width.3pt}  \hrule height.3pt}}}
\def\square{\mathchoice{\sqr67\,}{\sqr67\,}\sqr{3}{3.5}\sqr{3}{3.5}}
\def\today{\ifcase\month\or
  January\or February\or March\or April\or May\or June\or July\or
  August\or September\or October\or November\or December\fi
  \space\number\day, \number\year}

\def\Bbb{\bf}

\def\be{\begin{equation}}
\def\ee{\end{equation}}
\def\ba{\begin{eqnarray}}
\def\ea{\end{eqnarray}}

\preprint{DAMTP-1998-133}

\title{Open and Closed Cosmological Solutions of Ho\v{r}ava-Witten Theory}

\author{Harvey S. Reall \thanks{email: H.S.Reall@damtp.cam.ac.uk}}

\address {\qquad \\ DAMTP\\
Silver Street\\
Cambridge, CB3 9EW, UK
}

\maketitle

\begin{abstract}

The cosmological solutions of Ho\v{r}ava-Witten theory discovered by
Lukas, Ovrut and Waldram are generalized to allow non vanishing
spatial curvature. The solution with closed spatial sections has
initial and final five dimensional curvature singularities. 
We find two solutions with
open spatial sections, both of which evolve from an initial curvature
singularity to the supersymmetric domain wall solution at late times.
We also present a solution with open spatial sections and a non-zero
Ramond-Ramond scalar. The behaviour of the solutions in eleven
dimensions is discussed. 

\end{abstract}


\section{Introduction}

The strongly coupled limit of the $E_8\times E_8$ heterotic
superstring theory has been identified by Ho\v{r}ava and Witten with
$M-$theory compactified on a $S^1/Z_2$ orbifold with $E_8$ gauge
fields on each orbifold fixed plane \cite{hw1,hw2}. This can be
further compactified to four dimensions using a Calabi-Yau manifold
\cite{witten}. Matching the predicted values for the four dimensional
gravitational and GUT couplings leads one to the conclusion that the
orbifold is an order of magnitude larger than the Calabi-Yau space
\cite{witten,banks}. This has the interesting consequence that the
early universe may have been effectively five dimensional.

Cosmological solutions of Ho\v{r}ava-Witten theory have been
constructed from brane solutions in \cite{benakli1,benakli2}. These
have non-trivial gauge field configurations on the orbifold fixed
planes. An alternative approach has been used by Lukas \emph{et al}
who have constructed an effective five dimensional version of
Ho\v{r}ava-Witten theory compactified on a Calabi-Yau space and
shown that the theory admits a supersymmetric solution in which our spacetime
is identified with a four dimensional domain wall \cite{lukas1}. They
generalized this solutions to allow a cosmological time dependence by
seeking separable solutions of the equations of motion of their
effective theory \cite{lukas2}. These solutions correspond to making
the moduli of the domain wall solution time dependent and either
evolve from or to a five dimensional curvature singularity.

In this paper separable solutions with non-vanishing spatial curvature
are presented. We find a solution with closed spatial sections
and two solutions with open spatial sections. The latter solutions are
of particular interest because they approach the static domain wall
solution at late times. All of the solutions have an initial five
dimensional curvature singularity and the closed solution has a final
curvature singularity. These singularities are not resolved in the
eleven dimensional low energy theory. 

Solutions with a non vanishing `Ramond-Ramond' scalar (so called
because it would be a type II RR scalar if the orbifold of
Ho\v{r}ava-Witten theory were replaced by a circle) were found in
\cite{lukas2}. We have found such a solution with open spatial
sections. However it is of rather a special type and more general
solutions can be expected to exist. 

The first part of this paper consists of a short review of the five
dimensional effective theory constructed in \cite{lukas1}. The second
section together with the appendix gives an exhaustive discussion of
the various cases arising in the separation of variables. We present
it here because the details were omitted in \cite{lukas2}. In
particular we show that all separation constants must vanish, which
was assumed in \cite{lukas2}. The general solution of the separated
equations for the case of vanishing RR scalar but non-vanishing
spatial curvature is derived and a special solution with
non-vanishing RR scalar and open spatial sections is presented. The
final section discusses the behaviour of these solutions in five and
eleven dimensions and how our work relates to previous work on
cosmological solutions with varying moduli.

\section{five dimensional effective action}

Lukas \emph{et al} reduce Ho\v{r}ava-Witten theory to five dimensions
using the metric ansatz
\be
 ds^2=V^{-\frac{2}{3}}g_{\mu\nu}dx^{\mu}dx^{\nu}+V^{\frac{1}{3}}\Omega_{mn}dy^mdy^n.
\ee
Here $x^{\mu}$ ($0\le\mu\le 4$) are coordinate on the five dimensional
spacetime (which includes the orbifold direction) with metric $g_{\mu\nu}(x)$
and $y^m$ are coordinate on the Calabi-Yau space with metric
$\Omega_{mn}(y)$. $V(x)$ is a scalar field measuring the deformation
of the Calabi-Yau space. In performing the reduction it is necessary
to retain a non-zero mode of the three form potential of eleven
dimensional supergravity. See \cite{lukas1} for the details. For us the
relevant part of the reduced action is
\ba
 S=\int_{M_5}\sqrt{-g}\left(\frac{1}{2}R-\frac{1}{2}(\partial\phi)^2\right.&-&\left. e^{-\sqrt{2}\phi}\partial\xi\cdot\partial\bar{\xi}-\frac{1}{6}\alpha^2e^{-2\sqrt{2}\phi}\right)+{}\nonumber\\{}&+& \sqrt{2}\int_{M_4^{(1)}}\sqrt{-\tilde{g}}\alpha e^{-\sqrt{2}\phi}-\sqrt{2}\int_{M_4^{(2)}}\sqrt{-\tilde{g}}\alpha e^{-\sqrt{2}\phi},
\ea
where $\phi$ is defined by $V=e^{\sqrt{2}\phi}$, $\alpha$ is a
constant, $M_5$ is the five
dimensional spacetime bounded by the orbifold fixed planes $M_4^{(i)}$
and $\tilde{g}_{ij}$ ($0\le i,j \le 3$) denotes the pull-back of the
metric on $M_5$ onto $M_4^{(i)}$. $\xi$ is a scalar field related
to the components of the three form on the Calabi-Yau space.
We have omitted several terms from
the action; this is legitimate provided the current $j_{\mu}\equiv
i(\xi\partial_{\mu}\bar{\xi}-\bar{\xi}\partial_{\mu}\xi)$ 
vanishes. Let $\xi=e^{\rho+i\theta}$. Then $j_{\mu}$ vanishes if, and
only if, $\theta$ is constant. This is what we shall assume in the following.

Let $y\equiv x^4$ be a coordinate in the orbifold direction with $y\in
[-\pi\lambda,\pi\lambda]$ and $Z_2$ acting on $S^1$ by $y\rightarrow
-y$. The orbifold fixed planes are at $y=0,\pi\lambda$.
Then the equations of motion following from the action are
\ba
\label{eqn:einstein}
 G_{\mu\nu}=\partial_{\mu}\phi\partial_{\nu}\phi+2e^{2\rho-\sqrt{2}\phi}\partial_{\mu}\rho\partial_{\nu}\rho &-& g_{\mu\nu}\left(\frac{1}{2}(\partial\phi)^2+e^{2\rho-\sqrt{2}\phi}(\partial\rho)^2+\frac{1}{6}\alpha^2 e^{-2\sqrt{2}\phi}\right)+{}\nonumber\\{}&+&\sqrt{2}\alpha\sqrt{\frac{\tilde{g}}{g}}\tilde{g}^{ij}g_{i\mu}g_{j\nu}e^{-\sqrt{2}\phi}(\delta(y)-\delta(y-\pi\lambda)),
\ea
\be
 \frac{1}{\sqrt{-g}}\partial_{\mu}(\sqrt{-g}\partial^{\mu}\phi)=-\sqrt{2}e^{2\rho-\sqrt{2}\phi}(\partial\rho)^2-\frac{\sqrt{2}}{3}\alpha^2
e^{-2\sqrt{2}\phi}+2\alpha\sqrt{\frac{\tilde{g}}{g}}e^{-\sqrt{2}\phi}(\delta(y)-\delta(y-\pi\lambda))
\ee
\be
 \frac{1}{\sqrt{-g}}\partial_{\mu}(\sqrt{-g}e^{\rho-\sqrt{2}\phi}\partial^{\mu}\rho)=0,
\ee
where $G_{\mu\nu}$ is the five dimensional Einstein tensor and
$\tilde{g}^{ij}$ denotes the inverse of $\tilde{g}_{ij}$.

\section{separable cosmological solutions}

Following Lukas \emph{et al} we seek solutions of the equations of
motion of the form
\be 
 ds^2=-e^{2U(t,y)}dt^2+e^{2A(t,y)}ds_3^2+e^{2B(t,y)}dy^2,\qquad\phi=\phi(t,y),\qquad\rho=\rho(t,y),
\ee
where $ds_3^2$ is the line element on a three dimensional space of
constant curvature of sign $k\in\{-1,0,1\}$. Only the case
$k=0$ was considered in \cite{lukas2}. Substituting into the equations
of motion gives six partial differential equations. We shall solve
these by separation of variables. Set $U(t,y)=U_1(t)+U_2(y)$ etc. Then
the $tt$ component of the Einstein equation is
\ba
 e^{2(U_2-B_2)}(3A_2''+
6{A_2'}^2-3A_2'B_2'&+&\frac{1}{2}{\phi_2'}^2+e^{2\rho_1-\sqrt{2}\phi_1}e^{2\rho_2-\sqrt{2}\phi_2}{\rho_2'}^2)+{}\nonumber\\{}
+\frac{1}{6}\alpha^2
e^{2(B_1-\sqrt{2}\phi_1)}e^{2(U_2-\sqrt{2}\phi_2)}&-&\sqrt{2}\alpha
e^{B_1-\sqrt{2}\phi_1}e^{2U_2-B_2-\sqrt{2}\phi_2}(\delta(y)-\delta(y-\pi\lambda))\nonumber\\=e^{-2(U_1-B_1)}\left[3\dot{A_1}^2+3\dot{A_1}\dot{B_1}\right.&-&\left.\frac{1}{2}\dot{\phi_1}^2-e^{2\rho_1-\sqrt{2}\phi_1}e^{2\rho_2-\sqrt{2}\phi_2}\dot{\rho_1}^2\right]+{}\nonumber\\{}&+&3ke^{-2(A_1-B_1)}e^{-2(A_2-U_2)}.
\ea
Primes and dots denote derivatives with respect to $y$ and $t$
respectively. The delta functions are $y-$dependent so the terms
involving them must be made independent of $t$ if this equation is to
separate. Thus we need $B_1=\sqrt{2}\phi_1$. The final (curvature)
term can be made independent of either $y$ or $t$. We shall make the
first choice by setting $A_2=U_2$ (the latter choice would give the
scale factors of the three space and orbifold the same time
dependence; this is clearly not a good description of our universe).
Note that any constants arising in these relations can be absorbed
into $B_2$ or $A_1$. 
Finally we must deal with the $\rho$ terms. To separate these we must
either take $\rho=\sqrt{2}\phi+\textrm{constant}$, which is discussed
in the appendix, or take
\be
\label{eqn:xia}
 e^{-2(U_1-B_1)}e^{2\rho_1-\sqrt{2}\phi_1}\dot{\rho_1}^2=a^2,
\ee
\be
\label{eqn:xib}
 e^{2(U_2-B_2)}e^{2\rho_2-\sqrt{2}\phi_2}{\rho_2'}^2=b^2,
\ee
where $a$ and $b$ are constants. 

These conditions also
ensure the separation of the other equations of motion. After
separation if one adds the $y-$dependent part of the $yy$ equation to
the $y-$dependent parts of the $tt$ and $ij$ equations then the
$y-$dependent equations of motion can be written 
\be
\label{eqn:y1}
 \frac{d}{dy}\left(e^{4A_2-B_2}A_2'\right)+\frac{1}{9}\alpha^2e^{4A_2+B_2-2\sqrt{2}\phi_2}=\frac{1}{3}(\lambda_1+\lambda_3)e^{2A_2+B_2},
\ee
\be
\label{eqn:y2}
 \frac{d}{dy}\left(e^{4A_2-B_2}A_2'\right)+\frac{1}{9}\alpha^2e^{4A_2+B_2-2\sqrt{2}\phi_2}-\frac{2}{3}a^2e^{2A_2+B_2+2\rho_2-\sqrt{2}\phi_2}=\frac{1}{3}(\lambda_2+\lambda_3)e^{2A_2+B_2},
\ee
\be
\label{eqn:y3}
 6{A_2'}^2-\frac{1}{2}{\phi_2'}^2+\frac{1}{6}\alpha^2e^{2(B_2-\sqrt{2}\phi_2)}=\lambda_3e^{-2(A_2-B_2)},
\ee
\be
\label{eqn:y4}
 \frac{d}{dy}\left(e^{4A_2-B_2}\phi_2'\right)+\frac{\sqrt{2}}{3}\alpha^2e^{4A_2+B_2-2\sqrt{2}\phi_2}-\sqrt{2}a^2e^{2A_2+B_2+2\rho_2-\sqrt{2}\phi_2}=\lambda_4e^{2A_2+B_2}.
\ee
The $\lambda_i$ are the separation constants. 


If one adds the $t-$dependent part of the $tt$ equation to 
the $t-$dependent parts of the $ij$ and $yy$ equations then the
$t-$dependent equations of motion can be written
\be
\label{eqn:t1}
 3\dot{A_1}^2+3\dot{A_1}\dot{B_1}-\frac{1}{4}\dot{B_1}^2-b^2e^{2U_1-3B_1+2\rho_1}+3ke^{2(U_1-A_1)}=\lambda_1 e^{2(U_1-B_1)},
\ee
\be
\label{eqn:t2}
 \frac{d}{dt}\left(e^{-U_1+3A_1+B_1}(2\dot{A_1}+\dot{B_1})\right)-2b^2e^{U_1+3A_1-2B_1+2\rho_1}+4ke^{U_1+A_1+B_1}=(\lambda_1+\lambda_2)e^{U_1+3A_1-B_1},
\ee
\be
\label{eqn:t3}
 \frac{d}{dt}\left(e^{-U_1+3A_1+B_1}\dot{A_1}\right)+2ke^{U_1+A_1+B_1}=\frac{1}{3}(\lambda_1+\lambda_3)e^{U_1+3A_1-B_1},
\ee
\be
\label{eqn:t4}
 \frac{d}{dt}\left(e^{-U_1+3A_1+B_1}\dot{B_1}\right)-2b^2e^{U_1+3A_1-2B_1+2\rho_1}=\sqrt{2}\lambda_4e^{U_1+3A_1-B_1}.
\ee
Clearly the last three equations require
$3\sqrt{2}\lambda_4=\lambda_1+3\lambda_2-2\lambda_3$ for consistency.

The $ty$ equation of motion is 
\be
\label{eqn:ty}
 \left[e^{-U_1+\frac{3}{2}B_1-\rho_1}\dot{B_1}\right]\left[e^{A_2-B_2+\frac{1}{\sqrt{2}}\phi_2-\rho_2}(3A_2'-\frac{1}{\sqrt{2}}\phi_2')\right]=2ab.
\ee

The $\xi$ equation of motion separates to give
\be
\label{eqn:xiy}
 be^{A_2-B_2+\frac{1}{\sqrt{2}}\phi_2-\rho_2}(3A_2'-\frac{1}{\sqrt{2}}\phi_2')=\lambda_5,
\ee
\be
\label{eqn:xit}
 ae^{-U_1+\frac{3}{2}B_1-\rho_1}(3\dot{A_1}-\frac{1}{2}\dot{B_1})=\lambda_5.
\ee

We have not included the
delta function terms in the above equations so they are only valid in
$0<|y|<\pi\lambda$. The delta functions impose the following conditions on
the discontinuities of the derivatives of $A_2$ and $\phi_2$ at the
orbifold fixed planes
\be
 [A_2']=\pm\frac{\sqrt{2}}{3}\alpha e^{B_2-\sqrt{2}\phi_2},
\ee
\be
 [\phi_2']=\pm 2\alpha e^{B_2-\sqrt{2}\phi_2},
\ee
the upper sign refers to $y=0$ and the lower to $y=\pi\lambda$. The right
hand side is evaluated at the fixed plane. If one
combines these with the orbifold conditions $A(t,y)=A(t,-y)$ and
$\phi(t,y)=\phi(t,-y)$ then one obtains the following boundary
conditions at $y=0$
\be
\label{eqn:Abc}
 A_2'(\pm 0)=\pm\frac{\sqrt{2}}{6}\alpha e^{B_2-\sqrt{2}\phi_2},
\ee
\be
\label{eqn:phibc}
 \phi_2'(\pm 0)=\pm\alpha e^{B_2-\sqrt{2}\phi_2},
\ee
with similar conditions (but opposite signs) at $y=\pi\lambda$.
Substituting these boundary conditions into \ref{eqn:y3} and
\ref{eqn:xiy} gives $\lambda_3=\lambda_5=0$. 

If one subtracts \ref{eqn:y2} from \ref{eqn:y1} then one obtains
\be
\label{eqn:aeq}
 2a^2e^{2\rho_2-\sqrt{2}\phi_2}=\lambda_1-\lambda_2,
\ee
Equations
\ref{eqn:ty}, \ref{eqn:xiy}, \ref{eqn:xit} and \ref{eqn:aeq} can now
be satisfied
by choosing $a=0$ and $\phi_2'=3\sqrt{2}A_2'$. (There are other ways
of satisfying these equations but they lead to no new solutions. See
the appendix for details.)
Equation \ref{eqn:aeq} implies $\lambda_1=\lambda_2=0$ hence all
separation constants vanish. The vanishing of the separation constants was
assumed in \cite{lukas2}; we have shown that it is necessary for
the consistency of the equations and boundary conditions.
We are free to choose the
gauge $B_2=4A_2$ by redefining $y\rightarrow \tilde{y}(y)$. The
$y$-equations reduce to
\be
\label{eqn:ysub1}
 A_2''+\frac{1}{9}\alpha^2 e^{-4A_2}=0,
\ee
\be
\label{eqn:ysub2}
 {A_2'}^2-\frac{1}{18}\alpha^2 e^{-4A_2}=0.
\ee
It is then
straightforward to solve \ref{eqn:ysub2}. The result also satisfies
\ref{eqn:ysub1} (which can be obtained by differentiating \ref{eqn:ysub2}). 
The solution of the $y$-equations is the one found in
\cite{lukas1,lukas2}, namely
\be
\label{eqn:ydep}
 e^{A_2}=e^{U_2}=a_0 H^{\frac{1}{2}}, \qquad e^{B_2}=b_0 H^2,
\ee
where $H(y)=\frac{\sqrt{2}}{3}\alpha |y|+c_0$ and $a_0$, $b_0$ and
$c_0$ are constants. 

We are taking $a=0$ so \ref{eqn:xia} implies $\rho_1$ is a constant
that can be absorbed into $b$. The solution for $\xi$ is easily
obtained from \ref{eqn:xib}:
\be
 \xi=e^{i\theta}(d_0H^4+\xi_0)
\ee
where $\theta$, $d_0$ and $\xi_0$ are real constants. This solution of the
$y-$equations is the one found in \cite{lukas2}. 

Now consider the $t$-dependent equations. In the $k=0$ case considered
in \cite{lukas2} these were integrated by choosing a gauge so that the
`potential' (i.e. non-derivative) terms in \ref{eqn:t2}, \ref{eqn:t3}
and \ref{eqn:t4} become constant i.e. the gauge $U_1=-3A_1+2B_1$ was
used. This procedure does not work if both $b$ and $k$ are non-zero
because not all such terms can be made constant
simultaneously. However the equations can be cast in a form more
amenable to qualitative analysis by working in the four dimensional
Einstein frame with metric $\bar{g}_{ij}=e^Bg_{ij}$ i.e. the line
element is
\be
\label{eqn:fourd}
 ds_4^2=e^{2A_2+B_2}(-e^{2U_1+B_1}dt^2+e^{2A_1+B_1}ds_3^2).
\ee
Note that there is an overall $y-$dependent conformal factor so this
cannot really be interpreted as the metric of a four dimensional
spacetime. However this is irrelevant in what follows. Define a scale
factor $R(t)=e^{A_1+\frac{1}{2}B_1}$. Then in the
`comoving' gauge given by $U_1=-\frac{1}{2}B_1$ the $t-$dependent
equations reduce to
\be
\label{eqn:econs}
 \left(\frac{\dot{R}}{R}\right)^2=\frac{\kappa^2}{3}\left(\frac{1}{2}\dot{B_1}^2+V\right)-\frac{k}{R^2},
\ee
\be
\label{eqn:scalareq}
 \frac{1}{R^3}\frac{d}{dt}\left(R^3\dot{B_1}\right)=-\frac{dV}{dB_1},
\ee
where $V=\frac{1}{2}b^2e^{-2\sqrt{2}\kappa B_1}$ and $\kappa^2=8\pi G$
is the four dimensional Planck scale (which took the value 2 in the
units used previously). The other two $t-$equations are implied by
these two. These are the equations governing
the homogeneous and isotropic cosmology of a single scalar field $B_1$
with potential $V$. Note that $V$ is of the exponential type
considered in \cite{hawking} and that the factor of $2\sqrt{2}$
occuring in the exponent exceed the critical value of $\sqrt{2}$
required for a significant inflationary period. Exact analytic
solutions of \ref{eqn:econs} and \ref{eqn:scalareq} are only available
in the case $k=0$ considered in \cite{lukas2}. However special
solutions can be found by assuming that the curvature and potential
terms are proportional. Consider the slightly more general potential
$V_0e^{\alpha\kappa B_1}$. If one seeks a solution with $R(t)=R_0
e^{-\frac{1}{2}\alpha\kappa B_1}$ then one finds $R\propto t$
and $e^{\alpha\kappa B_1}\propto \frac{1}{t^2}$ subject to the
restriction
\be
 \frac{1}{4}\kappa^2 V_0 (2-\alpha^2)=\frac{k}{R_0^2},
\ee
so the geometry of the spatial sections in this solution is dictated
by the sign of $2-\alpha^2$. In our case we have $\alpha=-2\sqrt{2}$
and hence $k=-1$. The scale factor is $R=\frac{2}{\sqrt{3}}t$ which is
non-inflationary, as expected. 

Lukas \emph{et al} derived the general solution to the $t-$equations
in the $k=0$, $b\ne 0$ case. We shall now do the same for the $b=0$,
$k\ne 0$ case. The equations can be integrated in
the gauge given by $U_1=A_1$ i.e. using conformal time. Let
$Y=A_1+\frac{1}{2}B_1$. The $t$-equations are (using $\kappa^2=2$ again)
\be
\label{eqn:x1} 
 \dot{Y}^2-\frac{1}{4}\dot{B_1}^2-\frac{1}{6}\dot{\phi_1}^2+k=0,
\ee
\be
\label{eqn:ddY}
 \frac{d^2}{dt^2}e^{2Y}+4ke^{2Y}=0,
\ee
\be
 \frac{d}{dt}(e^{2Y}\dot{B_1})=0,
\ee
\be
 \frac{d}{dt}(e^{2Y}\dot{\phi_1})=0.
\ee
Note that we have reintroduced $\phi_1$ as an independent field. This
is so that we can compare our solutions of Ho\v{r}ava-Witten theory
with solutions of eleven dimensional supergravity compactifed on a
circle rather than an orbifold. If one compactifies from eleven to
five dimensions on a Calabi-Yau space and does not include a
non-zero mode of the three form on the internal space then the five
dimensional effective theory simply consists of gravity and a massless
scalar field $\phi_1$. One can seek cosmological solutions of this theory in
which the fifth dimension is assumed to be a circle, rather than the
orbifold considered above (in other words we are considering
cosmological solutions of type IIA supergravity compactified on a
Calabi-Yau space with $B_1$ the dilaton). 
There is then no need for the metric to
depend upon the circle direction so one does not have to separate
variables. The equations of motion are then the ones that we have just
written down. Of course to recover the solutions of Ho\v{r}ava-Witten
theory, we must set $\sqrt{2}\phi_1=B_1$ and 
include the $y-$dependence found above.

Solving the final two equations gives
\be
\label{eqn:B1dotsoln}
 \dot{B_1}=\beta e^{-2Y},\qquad \dot{\phi_1}=\gamma e^{-2Y},
\ee
where $\beta$ and $\gamma$ are constants. The other equations can
now be solved to give $e^{2Y}$ which can be substituted into
\ref{eqn:B1dotsoln}. Integration then yields the solutions
\be
 e^{A_1}=a_0 |\sin t|^{\frac{1-\delta}{2}}(\cos
t)^{\frac{1+\delta}{2}}\qquad e^{B_1}=b_0 |\tan t|^{\delta}\qquad
\phi_1=\textrm{const}+\epsilon\delta\log |\tan t|,
\ee
\be
 e^{A_1}=a_0 |t|^{\frac{1-\delta}{2}}\qquad e^{B_1}=b_0
|t|^{\delta}\qquad
\phi_1=\textrm{const}+\epsilon\delta\log |t|,
\ee
\be
 e^{A_1}=a_0 |\sinh t|^{\frac{1-\delta}{2}}(\cosh
t)^{\frac{1+\delta}{2}}\qquad e^{B_1}=b_0 |\tanh t|^{\delta}\qquad
\phi_1=\textrm{const}+\epsilon\delta\log |\tanh t|,
\ee
in the cases $k=+1,0,-1$ respectively. The constants $\epsilon$ and
$\delta$ are defined by
\be
 \epsilon=\frac{\gamma}{\beta}\qquad\delta=\frac{\beta}{\sqrt{\beta^2+\frac{2\gamma^2}{3}}}.
\ee
We see that there is one parameter family of solutions of the type IIA
theory. Note that our solutions of Ho\v{r}ava-Witten theory require
$\epsilon=\frac{1}{\sqrt{2}}$ so $\delta=\pm\frac{\sqrt{3}}{2}$.

If we define
\be
   {\tau=\left\{\begin{array}{ll}
         \tan t & k=+1 \\
	 t & k=0 \\
	 \tanh t & k=-1
         \end{array}\right.}
\ee
then the solutions can be written in the unified form
\be
   e^{A_1}=\frac{|\tau|^{\frac{1-\delta}{2}}}{\sqrt{1+k\tau^2}}\qquad
e^{B_1}=|\tau|^{\delta}\qquad \phi_1
=\textrm{const}+\epsilon\delta\log|\tau|.
\ee 

To summarize, in Ho\v{r}ava-Witten theory our separable cosmological
solutions have the $y-$dependence of the static domain wall given by
\ref{eqn:ydep} and time dependence given by the above results with
$\delta=\pm\frac{\sqrt{3}}{2}$. Thus, for example, in the $k=-1$ case
the metric is
\be
 ds^2=a_0^2 |\sinh t|^{1\mp\frac{\sqrt{3}}{2}}(\cosh
t)^{1\pm\frac{\sqrt{3}}{2}}H(y)(-dt^2+ds_3^2) + b_0^2 |\tanh
t|^{\pm\sqrt{3}}H(y)^4 dy^2,
\ee
and the scalar field is
\be
 V\equiv e^{\sqrt{2}\phi}=b_0 |\tanh t|^{\pm\frac{\sqrt{3}}{2}}H(y)^3,
\ee
where $a_0$ and $b_0$ are positive constants.

\section{discussion of solutions}

We have seen that cosmological solutions of Ho\v{r}ava-Witten theory
compactified on a Calabi-Yau space can be obtained by separation of
variables. Solutions with curved spatial sections can be obtained and
their time dependence is seen to correspond to particular members of a
family of type IIA cosmological solutions. 

We shall first discuss the solutions with vanishing Ramond-Ramond scalar.
The $k=0$ solutions derived above are the same as those derived in
\cite{lukas2}. This can be seen by changing to comoving time, $T$,
defined by $dT=e^{A_1} dt$. However our $k=\pm 1$ solutions are new. At
early times these solutions behave just like the $k=0$ ones. This is
exactly what happens in conventional (four dimensional) cosmology:
spatial curvature is negligible in the early universe. Here we have
recovered this result in a five dimensional setting.

The range of $t$ can be divided into two regions, namely $t<0$ and
$t>0$. In \cite{lukas2} these were referred to as the $(-)$ and $(+)$
branches respectively and are related by time reversal. We shall only
consider the $(+)$ solutions (corresponding to a universe that is
initially expanding).

The five dimensional Ricci scalar (computed by taking the trace of
\ref{eqn:einstein}) diverges at $t=0$ for all of the above
solutions, indicating the presence of a five dimensional curvature
singularity. However this may be merely a singularity of the effective
theory that gets resolved in the full eleven dimensional theory. It is
therefore important to examine the solutions from an eleven
dimensional perspective. In the $k=0$ case the metric is
\be
 ds^2=a_0^2t^{1\mp\frac{5}{6}\sqrt{3}}H(y)^{-1}\eta_{ij}dx^idx^j+b_0^2t^{\pm\frac{2}{3}\sqrt{3}}H(y)^2dy^2 + c_0^2t^{\pm\frac{1}{6}\sqrt{3}}H(y)ds_6^2,
\ee
where $a_0$, $b_0$ and $c_0$ are constants. 
We shall refer to the upper and lower choices of sign as the $(\uparrow)$ and
$(\downarrow)$ solutions respectively. The qualitative behaviour of
these solutions near $t=0$ was discussed in \cite{lukas2} in the five
dimensional theory. In eleven dimensions the qualitative behaviour of
the Calabi-Yau space is the same as that of the orbifold
(i.e. collapsing for the $(\uparrow)$ solutions and decompactifying
for the $(\downarrow)$ solutions).  
It is straightforward to compute the
eleven dimensional Ricci scalar, which is found to diverge at $t=0$
for the $(\uparrow)$ solution and vanish there for the $(\downarrow)$ 
solution. Thus
the $(\uparrow)$ solution has a genuine singularity at $t=0$. For 
the $(\downarrow)$
solution, the square of the eleven dimensional Ricci tensor vanishes
at $t=0$ but the square of the Riemann tensor is divergent there (the
formulae in \cite{ivashchuk} are useful for computing these quantities). Hence
$t=0$ is also a singularity in this case but of a rather different
type than in the $(\uparrow)$ solution. For both types of solution the
singularity lies at a finite affine parameter distance in the past on
timelike and null geodesics. 

After the initial singularity, the induced metrics on the (four
dimensional) orbifold fixed planes evolve as non-inflationary FRW
universes. For example in the $k=0$ case \cite{lukas2} the scale
factor is $T^{\frac{3}{11}(1\mp \frac{4}{3\sqrt{3}})}$.

Reducing the bulk metric to four dimensions involves taking account of
the massive modes arising through integrating out the
$y$-dependence. However the time dependence of the resulting four
dimensional Einstein frame metric can be read off from
\ref{eqn:fourd}. Interestingly, the $(\uparrow)$ and $(\downarrow)$
solutions both give the same result after this reduction. In the $k=0$
case (the only case for which a simple analytic expression exists),
the scale factor is $T^{\frac{1}{3}}$. This degeneracy also arises in
the reduction of the type IIA solutions discussed above: the four
dimensional Einstein frame metric is independent of $\delta$. 

At late times the behaviour of the solutions is sensitive to the
spatial curvature. The $k=0$ solutions undergo power law expansion or
contraction as discussed in \cite{lukas2}. The $k=+1$ solutions have
interesting behaviour near $t=\frac{\pi}{2}$. Before analyzing this we
should first mention that in the $k=+1$ case there is only one solution
because the $(\uparrow)$ and $(\downarrow)$ metrics are related by 
the coordinate transformation $t\rightarrow \frac{\pi}{2}-t$. We 
shall consider the metric in the $(\uparrow)$ form. In the five dimensional
theory the three dimensional spatial sections
collapse to zero size at $t=\frac{\pi}{2}$ and the orbifold
decompactifies. The five dimensional Ricci scalar diverges. 
The qualitative behaviour of the spatial section and orbifold is 
unchanged in the eleven dimensional metric and the behaviour of the 
Calabi-Yau space is
qualitatively the same as that of the orbifold. The eleven dimensional
Ricci scalar and Ricci tensor squared both vanish at $t=\frac{\pi}{2}$
(since we know that this is what happens for the $(\downarrow)$ form of the
metric at $t=0$) but the square of the Riemann tensor diverges. Thus
this solution expands from an initial curvature singularity to a
final one but the nature of the two singularities is different.

The most interesting case is $k=-1$. For large $t$ it has
$e^{A_1}\propto e^t$ and $e^{B_1}=\textrm{constant}$. By changing to
comoving time one immediately sees that this is nothing but the static
domain wall solution. Thus both the $(\uparrow)$ and $(\downarrow)$ 
solutions evolve 
from an initial curvature singularity to a supersymmetric solution
appropriate for the reduction to a $N=1$ supergravity theory in four
dimensions \cite{witten,lukas1}, which has phenomenological appeal. 

Finally we turn to the special $k=-1$ solution with non-vanishing RR
scalar. 
The scale factor is proportional to the comoving time.
In five dimensions the Ricci scalar diverges at $t=0$. The
eleven dimensional metric is
\be
 ds^2=a_0^2H(y)^{-1}\left(-dT^2+\frac{49}{108}T^2ds_3^2\right)+b_0^2T^{\frac{8}{7}}H(y)^2dy^2+c_0^2T^{\frac{2}{7}}H(y)ds_6^2,
\ee
where $T$ is the (eleven dimensional) comoving time. The eleven
dimensional Ricci scalar diverges at $T=0$. 
Note that the expansion at late times in this solution is slower 
than in the $k=-1$ solutions with no RR scalar.

We have seen that the cosmological solutions of Lukas \emph{et al} can
be easily generalized to include non-zero spatial curvature. The time
dependence is the same as that of particular solutions for a massless
scalar field evolving in five dimensions with the fifth dimension a
circle rather than an orbifold. Massless scalar fields typically arise
when one allows moduli to vary. These were considered in detail by
Gibbons and Townsend \cite{gibbons}. Our solutions share the generic
properties found by them. This may seem suprising because we have solutions
varying in two directions whereas theirs only varied in one. However
we have seen that separation of variables forces the potential term
into the $y-$equations so the time evolution is just that of a
massless scalar field. More general (i.e. non-separable) solutions can
be expected to have more interesting time dependence.

\medskip
\centerline {\bf Acknowledgements}

I have enjoyed useful conversations with Andrew Chamblin, Gary
Gibbons, Stephen Hawking and Neil Turok.

\appendix
\section*{}

This appendix deals with the cases encountered in the analysis of
section III that do not lead to new solutions.

We first discuss what would have happened had we taken
$\rho=\sqrt{2}\phi+\textrm{constant}$ in order to separate the field 
equations instead of \ref{eqn:xia} and \ref{eqn:xib}. Define a
constant $\beta$ by
\be
 \beta^2=2e^{2\rho-\sqrt{2}\phi}.
\ee
Then the separated equations of motion are the as those derived above
(with $a=b=0$) with the exceptions that \ref{eqn:y3} is replaced by
\be
 6{A_2'}^2-\frac{1}{2}(1+\beta^2){\phi_2'}^2+\frac{1}{6}\alpha^2e^{2(B_2-\sqrt{2}\phi_2)}=\lambda_3e^{-2(A_2-B_2)},
\ee
\ref{eqn:t1} is replaced by
\be
\label{eqn:y3alt}
 3\dot{A_1}^2+3\dot{A_1}\dot{B_1}-\frac{1}{4}(1+\beta^2)\dot{B_1}^2+3ke^{2(U_1-A_1)}=\lambda_1 e^{2(U_1-B_1)}
\ee
and \ref{eqn:ty} is replaced by
\be
 \dot{B_1}\left(3\sqrt{2}A_2'-(1+\beta^2)\phi_2'\right)=0.
\ee
The solution $(1+\beta^2)\phi_2'=3\sqrt{2}A_2'$ of the final equation
is incompatible with the orbifold boundary conditions unless $\beta=0$
i.e. $\rho=-\infty$ i.e. $\xi=0$. So this will simply lead to one of
the solutions derived above for the case $a=b=0$. The second
possibility is $\dot{B_1}=0$. The relations $\lambda_1=\lambda_2$ and
$3\sqrt{2}\lambda_4=\lambda_1+3\lambda_2-2\lambda_3$ are easily
derived as before and $\dot{B_1}=0$ implies $\lambda_4=0$ so
$\lambda_3=2\lambda_1$. If one now subtracts \ref{eqn:y4} from
\ref{eqn:y1}, chooses the gauge $B_2=-2A_2$ and integrates then one
obtains
\be
 e^{6A_2}(A_2'-\frac{1}{3\sqrt{2}}\phi_2')=\lambda_1(y-y_0),
\ee
where $y_0$ is a constant. This is clearly incompatible with the
boundary conditions unless $\lambda_1=0$. Hence all of the separation
constants vanish. Now substituting the boundary conditions into
\ref{eqn:y3alt} yields $\beta=0$ hence $\xi=0$. Thus this case does
not lead to any new solutions either. 

Now we turn to the alternative ways of satisfying equations
\ref{eqn:ty}, \ref{eqn:xiy}, \ref{eqn:xit} and \ref{eqn:aeq}. These
all requre either \emph{i}) $a=b=0$, $\dot{B_1}=0$; or 
\emph{ii}) $b=0$, $\rho_2=\frac{1}{\sqrt{2}}\phi_2$; 
Case \emph{i}) requires $\lambda_4=0$
hence (since $\lambda_3=0$) $\lambda_2=-\frac{1}{3}\lambda_1$. However
\ref{eqn:y1} and \ref{eqn:y2} require $\lambda_1=\lambda_2$. Hence all
separation constants must vanish. Equation \ref{eqn:t1} can only be
solved for $k=0,-1$. The solution is just the static domain wall.
For case \emph{ii}), equation \ref{eqn:xib} implies that
$\phi_2'=0$. However this is incompatible with the boundary
conditions so there are no solutions in this case.

\end{document}